\documentclass[letterpaper]{article} 
\usepackage{aaai2027}  
\usepackage[hyphens]{url}  
\usepackage{graphicx} 
\usepackage{natbib}  
\usepackage{caption} 
\usepackage{amsmath}
\usepackage{multirow}
\usepackage{amssymb}
\usepackage{bbding}
\usepackage{tcolorbox}
\tcbuselibrary{skins}

\usepackage{algorithm}
\usepackage{algorithmic}

\usepackage{newfloat}
\usepackage{listings}
\DeclareCaptionStyle{ruled}{labelfont=normalfont,labelsep=colon,strut=off} 
\floatstyle{ruled}
\newfloat{listing}{tb}{lst}{}
\floatname{listing}{Listing}

\usepackage{booktabs}

\title{HIERA: Workload-Aware Planning Across Implementation Spaces \\ for GPU Kernel Optimization}

\author{
Jinghao~Wang\textsuperscript{1,2},\,
Qiqi Gu\textsuperscript{1,2},\,
Chenpeng Wu\textsuperscript{1,2},\,
Jianguo~Yao\textsuperscript{1,2},\,
Haibing~Guan\textsuperscript{1,2},\,
Xijun~Li\textsuperscript{1,2,\Envelope}\\
}

\affiliations{
1 Shanghai Key Laboratory of Scalable Computing and Systems\\
2 School of Computer Science, Shanghai Jiao Tong University\\
\Envelope Corresponding author
}

\begin{document}

\maketitle

\begin{abstract}
High-performance GPU kernels underpin modern deep learning and scientific computing. As workloads become increasingly diverse and GPU hardware evolves rapidly, developing efficient methods for automated GPU kernel generation and optimization has become increasingly important. Existing LLM-based methods typically optimize within a fixed implementation space, limiting either optimization flexibility or search efficiency. We propose \textsc{HIERA}, a hierarchical search-space planning framework for GPU kernel optimization. \textsc{HIERA} constructs contract-augmented task specifications, selects an appropriate implementation space across PyTorch operators, CUDA libraries, and custom CUDA kernels, and uses profiling feedback and expert knowledge to guide structured iterative refinement. Experiments on KernelBench across multiple various workload levels and base LLMs show that \textsc{HIERA} delivers stronger overall implementation validity, sample efficiency, and optimization performance than existing training-free methods, while remaining competitive with the training-based CUDA-L1 without additional model training. A case study on a specialized stencil operator from scientific computing further achieves a \(1.53\times\) speedup over cuDNN, demonstrating the potentiality of the general framework beyond standard machine-learning workloads.
\end{abstract}



\section{Introduction}

High-performance GPU kernels form a critical foundation of modern deep learning and scientific computing systems~\cite{chen2018tvm,zheng2020ansor}. To simplify GPU programming, vendor libraries and framework primitives, such as cuBLAS, cuDNN, and PyTorch operators, provide optimized implementations for common computation patterns~\cite{nvidia2026cublas,chetlur2014cudnn,paszke2019pytorch}. By encapsulating widely used kernels and established optimization strategies, these abstractions substantially reduce the effort required for GPU kernel development and optimization~\cite{chetlur2014cudnn,paszke2019pytorch}. However, relying exclusively on these abstractions remains insufficient. Rapid hardware evolution makes optimized implementations costly to maintain and retune across architectures~\cite{chetlur2014cudnn,chen2018tvm,zheng2020ansor}, while fixed library primitives cannot adequately cover the growing diversity of specialized workloads, such as complex scientific-computing operators~\cite{tillet2019triton,shi2023welder,zhuang2024mononn,holewinski2012stencil}. As existing libraries and framework primitives increasingly struggle to meet evolving workload and hardware requirements, GPU kernel development and optimization still rely heavily on manual implementation and tuning. This process demands substantial domain expertise and engineering effort, motivating the development of automated approaches to GPU kernel generation and optimization~\cite{chetlur2014cudnn,ouyang2025kernelbench,zhang2025cudaforge}.

\begin{figure}[t]
\centering
\includegraphics[width=\columnwidth]{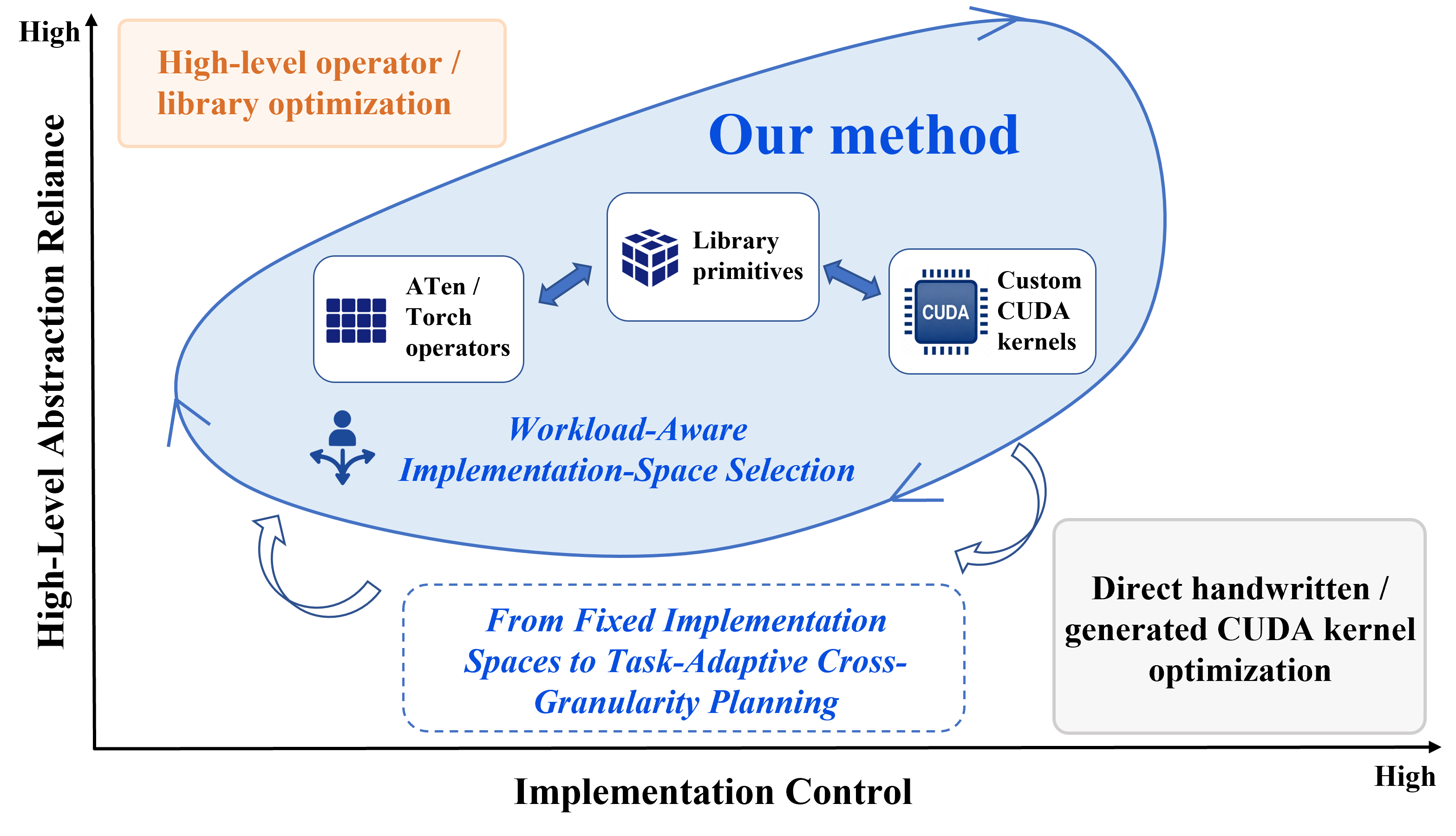}
\caption{
Comparison of prior methods operating within fixed implementation spaces and our proposed \textsc{HIERA}'s workload-aware cross-granularity planning.
}
\label{fig:overview}
\end{figure}

Recent LLM-based systems have automated GPU kernel generation through iterative code synthesis and execution-driven refinement~\cite{ouyang2025kernelbench,zhang2025cudaforge,wei2025astra,andrews2025gpukernelscientist}. Nevertheless, many existing methods optimize within a predetermined implementation space~\cite{zhang2025cudaforge,li2026cudal1improvingcudaoptimization,baronio2026kevin,wang2025geak,saba2026cutegen}. As illustrated in Figure~\ref{fig:overview}, methods concentrated in the upper-left region rely primarily on high-level operators or optimized libraries. These approaches generally preserve higher-level semantics and reduce implementation complexity~\cite{chen2018tvm,zheng2020ansor}, but may miss optimization opportunities that require custom fusion, shape specialization, or direct control over GPU execution~\cite{shi2023welder,zhuang2024mononn,cheng2026mpk}. In contrast, methods in the lower-right region directly explore custom CUDA implementations, providing greater optimization freedom but requiring the model to navigate a large and weakly structured space of interacting low-level decisions~\cite{zhang2025cudaforge,chen2025cudallm,li2026cudal1improvingcudaoptimization,baronio2026kevin}. Under limited candidate-generation budgets, such exploration can spend substantial resources on implementations that fail compilation or correctness checks, or yield little performance improvement~\cite{ouyang2025kernelbench,zhang2026kernelbenchverified,sarkar2026correctnessillusion,li2026correctbutslow}. More importantly, execution and profiling feedback can guide refinement only within the chosen implementation space; they do not determine whether that space is appropriate for the workload~\cite{zhang2025cudaforge,wei2025astra,andrews2025gpukernelscientist}. This limitation motivates \textit{treating implementation-space selection itself as an explicit part of the optimization process}.

To address this gap, we propose \textsc{HIERA}, a hierarchical search-space planning framework for GPU kernel optimization. \textsc{HIERA} treats implementation granularity as an explicit optimization decision: it first selects a workload-appropriate implementation space and then structures refinement within that space using profiling feedback and expert domain knowledge. The main contributions of this work are summarized as follows:

\begin{itemize}
\item We identify the limitations of predetermined implementation spaces for efficient LLM-based GPU kernel optimization, and formulate implementation-space selection as an explicit workload-aware planning problem.

\item We introduce a contract-augmented task specification that fixes callable interfaces, parameter semantics, compilation rules, and verification behavior. This design focuses candidate generation on performance-critical implementation choices and reduces ambiguity in code generation and evaluation.

\item We develop a hierarchical search-space planning mechanism that first selects an appropriate implementation regime spanning custom CUDA kernels, optimized CUDA libraries, and PyTorch operators, and then combines profiling feedback with expert-curated optimization knowledge to guide structured refinement within the selected space.

\item We conduct comprehensive comparative, limited-budget, and ablation studies on KernelBench across three workload levels and three base LLMs. The results show that \textsc{HIERA} consistently improves implementation validity, sample efficiency, and optimization performance over existing training-free methods, while confirming the complementary contributions of contract augmentation and hierarchical planning. A stencil case study further demonstrates its feasibility on a specialized scientific-computing operator under the evaluated configuration.

\end{itemize}

\section{Related Work}

\paragraph{Tensor Compilers and GPU DSLs.}
Tensor compilers and GPU programming systems provide structured optimization spaces for efficient kernel generation. TVM and Ansor search over compiler schedules and program transformations, while Triton provides a high-level language and compiler for custom GPU kernels~\cite{chen2018tvm,zheng2020ansor,tillet2019triton}. However, these systems remain constrained by predefined compiler IRs, scheduling languages, or DSL models~\cite{feng2023tensorir}. In contrast, \textsc{HIERA} treats the choice among high-level operators, optimized libraries, and custom CUDA implementations as part of the optimization process.

\paragraph{GPU Kernel Generation Benchmarks.}
KernelBench evaluates whether LLMs can generate GPU implementations that are both functionally correct and faster than their PyTorch references. Its original release contains 250 workloads across three levels, and adopts the $\mathrm{fast}_p$ metric to report the fraction of valid implementations exceeding a specified speedup threshold~\cite{ouyang2025kernelbench,kernelbench_repo}. We use KernelBench for evaluation, but augment its task artifacts with contract-augmented specifications that fix callable bindings, parameter semantics, and compilation rules, thereby focusing generation on contract-compliant implementations.

\paragraph{Agentic, Feedback-Driven CUDA Optimization.}
CUDAForge is a training-free Coder--Judge framework that iteratively refines CUDA implementations using correctness results, GPU specifications, and Nsight Compute metrics. It demonstrates the effectiveness of hardware-aware feedback for low-level optimization, but restricts exploration to a fixed custom-CUDA space~\cite{zhang2025cudaforge}. This restriction can be inefficient for composite and model-level workloads, where reconstructing operator dependencies and intermediate data flows directly in CUDA consumes substantial search resources. \textsc{HIERA} instead selects an implementation space according to workload characteristics before performing structured within-space refinement.

\paragraph{Reinforcement Learning for CUDA Optimization.}
CUDA-L1 uses contrastive reinforcement learning to favor correct and efficient CUDA implementations, while Kevin trains models on multi-turn refinement trajectories using correctness and speedup feedback~\cite{li2026cudal1improvingcudaoptimization,baronio2026kevin}. These methods encode optimization preferences implicitly in model parameters and require training. In contrast, \textsc{HIERA} uses explicit planning decisions that can be inspected and extended without retraining the underlying model.

Overall, prior work has advanced GPU kernel optimization through compiler abstractions, standardized benchmarks, feedback-driven refinement, and reinforcement learning. However, these approaches generally optimize within predefined representations or implementation spaces and do not explicitly treat implementation granularity as an optimization decision. \textsc{HIERA} addresses this gap by combining contract-augmented task specifications with workload-aware cross-granularity implementation-space planning and expert-guided optimization-direction pruning, enabling structured coarse-to-fine optimization across heterogeneous workloads.

\begin{figure*}[t]
    \centering
    \includegraphics[width=0.98\textwidth]{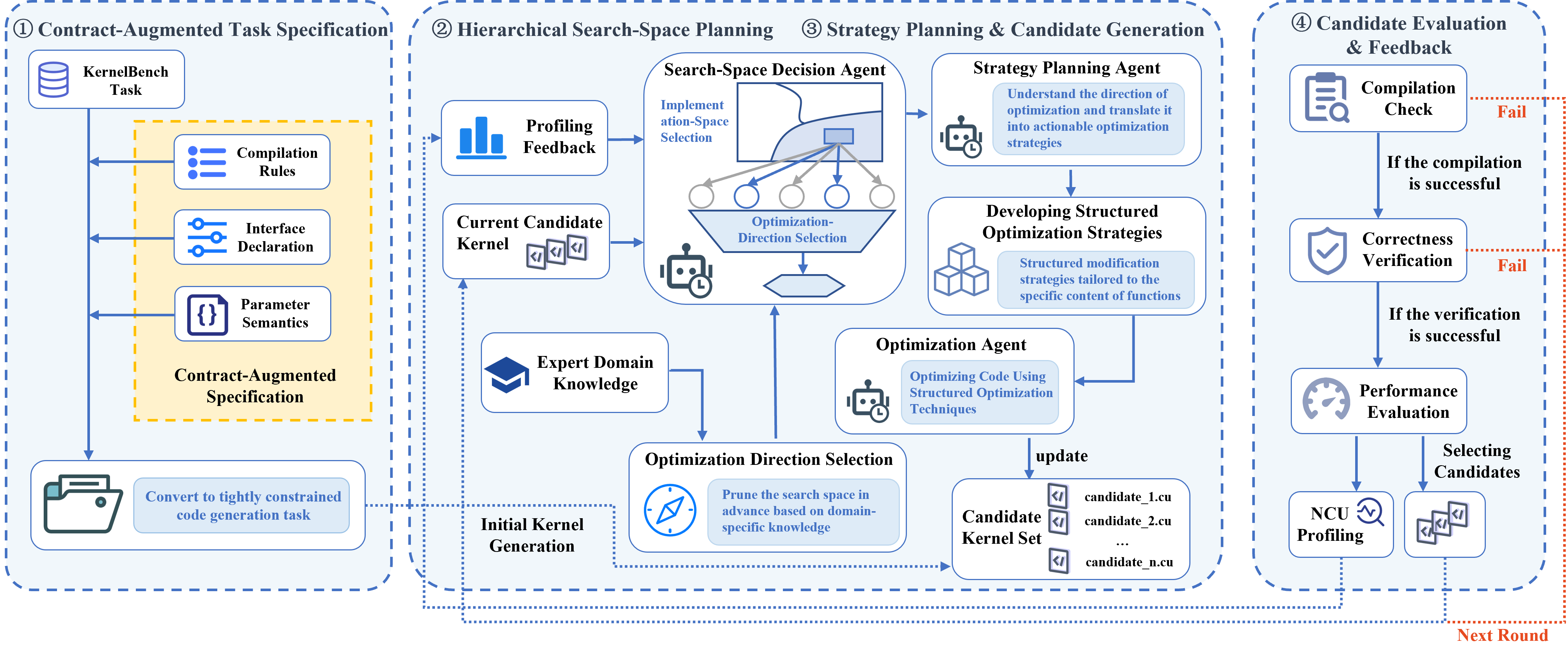}
    \caption{Overview of HIERA. The framework constructs a contract-augmented task specification, performs hierarchical planning over implementation spaces and optimization directions, and iteratively refines candidate implementations using evaluation and profiling feedback.}
    \label{fig:framework}
\end{figure*}

\section{Method}

\subsection{Problem Formulation and Framework Overview}

Given a kernel optimization task $\tau$, a reference implementation $f_{\tau}$, a target hardware platform $\mathcal{H}$, and a finite search budget $B$, our objective is to identify the fastest valid implementation explored within the budget. Let $\mathcal{C}_{B}(\tau)$ denote the set of explored candidate implementations, and let $\mathrm{Valid}(x;\tau) \in \{0,1\}$ indicate whether candidate $x$ compiles successfully and satisfies the interface and semantic requirements of $\tau$. For a valid candidate, its speedup over the reference implementation is defined as
\begin{equation}
s(x;\tau,\mathcal{H}) =
\frac{t(f_{\tau};\mathcal{H})}{t(x;\tau,\mathcal{H})},
\end{equation}
where $t(\cdot;\mathcal{H})$ denotes the execution latency on the target hardware. The optimization objective is
\begin{equation}
x_{\tau}^{*} =
\arg\max_{x \in \mathcal{C}_{B}(\tau)}
s(x;\tau,\mathcal{H})
\quad
\text{s.t.}
\quad
\mathrm{Valid}(x;\tau)=1.
\label{eq:optimization-objective}
\end{equation}

Figure~\ref{fig:framework} presents the overall workflow of HIERA. The framework first combines the KernelBench task with fixed template artifacts, interface declarations, parameter semantics, and compilation rules to construct a contract-augmented task specification. The Search-Space Decision Agent then performs hierarchical planning by selecting a workload-appropriate implementation space, followed by an optimization direction within that space based on the current candidate, profiling feedback, and expert domain knowledge. The Strategy Planning Agent translates the selected direction into structured optimization strategies, which guide the Optimization Agent in generating candidate implementations. The candidates undergo compilation checking, correctness verification, performance evaluation, and NCU profiling, after which selected candidates and their feedback are used in the next refinement round.

\subsection{Contract-Augmented Task Specification}

Generating a complete executable program unnecessarily expands the search space, as the model must reconstruct host-side wrappers, callable bindings, data-preparation logic, reference code, and build configurations that are fixed by the task and are not optimization targets. HIERA therefore converts each KernelBench task into a contract-augmented specification that explicitly defines the compilation rules, interface declarations, and parameter semantics. These contracts preserve the required external behavior while restricting generation to performance-critical implementation choices.

Concretely, for every KernelBench task, HIERA keeps the extension wrapper (\texttt{torch\_demo.py}), C++ binding source (\texttt{cpp\_source.cpp}), and correctness reference (\texttt{groundtruth.py}) fixed. Level~2 and Level~3 tasks additionally include a parameter-semantics file (\texttt{params\_semantics.json}) describing argument roles, constraints, and dependencies, whereas Level~1 tasks omit this file because their parameter semantics can be inferred from the interface and reference implementation. The model generates only the candidate implementation in \texttt{cuda\_source.cu} using a structured response. Each candidate is then combined with the fixed task artifacts and compiled through the prescribed PyTorch-extension entry point. This design eliminates repeated boilerplate generation and focuses the search budget on performance-relevant decisions.

\subsection{Cross-Granularity Search-Space Planning}

Most existing kernel optimization methods operate within a fixed implementation level or representation. However, the appropriate implementation granularity depends strongly on the workload~\cite{wu2025mirage}. For simple operator-level tasks, composing high-level operators may introduce dispatch and intermediate-materialization overheads~\cite{shi2023welder,zhuang2024mononn}. Conversely, implementing model-level workloads entirely in custom CUDA may require reconstructing complex operator dependencies, intermediate data flows, and execution logic, consuming substantial search budget before performance optimization begins. HIERA therefore treats implementation-space selection as an explicit planning decision.

At planning step $t$, the Search-Space Decision Agent (SSDA) first selects one of three nested implementation spaces with increasing permissiveness: \textit{pure CUDA}, \textit{CUDA libraries}, or \textit{CUDA libraries with PyTorch operators}. The first permits only custom CUDA kernels; the second additionally permits optimized libraries such as cuBLAS; and the third further permits high-level PyTorch operators. The selected space $g_t$ defines the admissible implementation choices for the current refinement step, avoiding unnecessary low-level reconstruction for composite workloads while preserving fine-grained control when custom CUDA optimization is advantageous.

\subsection{Domain-Guided Optimization-Direction Pruning}

Within the selected implementation space, the Search-Space Decision Agent (SSDA) further prunes the refinement space by selecting an optimization direction. Based on recurring bottlenecks and optimization principles identified in prior GPU optimization studies and expert practice, we summarize five broadly applicable directions that cover the major performance dimensions targeted by modern GPU kernels. This taxonomy is operational rather than exhaustive and can be extended with workload or architecture-specific directions.

Specifically, we consider \textit{control-flow and boundary specialization} ($C$), \textit{thread- and warp-level parallelism} ($P$), \textit{memory transaction efficiency} ($M$), \textit{data reuse and data-movement pipelining} ($R$), and \textit{Tensor Core and instruction-pipeline utilization} ($T$)~\cite{wang2025tilelangcomposabletiledprogramming,shi2023welder,weng2021unit,spector2024thunderkittens}. Let $\mathcal{D}=\{C,P,M,R,T\}$ denote the candidate direction set. These directions correspond to divergent or redundant execution, insufficient parallelism, inefficient global-memory transactions, limited on-chip reuse or data-movement overlap, and underutilized compute instructions, respectively.

At refinement step $t$, the agent scores each direction using the task specification $\tau$, selected implementation space $g_t$, current candidate $x_t$, profiling feedback $F_t$, expert knowledge $\mathcal{E}$, and a fixed scoring rubric $\mathcal{R}$:
\begin{equation}
\mathbf{s}_t =
\operatorname{SSDA}
\left(
\tau,g_t,x_t,F_t,\mathcal{E};\mathcal{R}
\right)
=
\left(s_t^d\right)_{d\in\mathcal{D}},
\label{eq:direction_scores}
\end{equation}
where $s_t^d$ denotes the relevance score of direction $d$. The primary direction is selected as
\begin{equation}
d_t^{*}
=
\arg\max_{d\in\mathcal{D}} s_t^d.
\label{eq:direction_selection}
\end{equation}

The Strategy Planning Agent (SPA) translates the selected direction into a structured, actionable optimization strategy:
\begin{equation}
\pi_t =
\operatorname{SPA}
\left(
\tau,g_t,x_t,F_t,d_t^{*},\mathcal{E}
\right),
\qquad
\pi_t \in \Pi_{g_t,d_t^{*}},
\label{eq:optimization_plan}
\end{equation}
where $\pi_t$ specifies concrete transformation steps and implementation constraints, and $\Pi_{g_t,d_t^{*}}$ denotes the strategy space admissible under $g_t$ and $d_t^{*}$. The Optimization Agent (OA) then applies this strategy to generate the next candidate kernel set:
\begin{equation}
\mathcal{X}_{t+1} =
\operatorname{OA}
\left(
\tau,g_t,x_t,F_t,\pi_t
\right).
\label{eq:optimization_generation}
\end{equation}

\subsection{Feedback-Driven Candidate Evaluation and Iteration}

Each generated candidate is combined with the fixed template code and subjected to compilation checking and correctness verification. Candidates that fail either check are discarded. Valid candidates are then evaluated on the target GPU and profiled with NVIDIA Nsight Compute (NCU) to obtain execution latency and hardware-level bottleneck metrics.

After each evaluation round, the framework records the candidate source code, compilation result, correctness outcome, execution latency, profiling output, selected implementation space, and optimization direction. Valid candidates are ranked according to measured performance, and the strongest candidates are retained as parents for the next iteration. The resulting feedback is provided to the decision and planning modules, allowing subsequent refinement to be guided by both observed performance and hardware-level evidence. The complete record of intermediate decisions and evaluation outcomes makes each optimization trajectory traceable and supports subsequent diagnosis and analysis.

\section{Experimental Setup}

\subsection{Research Questions}

\newtcolorbox{rqbox}[1][]{
    enhanced,
    colback=gray!3,
    colframe=black,
    coltitle=white,
    fonttitle=\bfseries,
    boxed title style={
        colback=black,
        sharp corners
    },
    attach boxed title to top left={
        xshift=3mm,
        yshift=-2.5mm
    },
    title={#1}
}

To systematically evaluate the effectiveness and behavior of
\textsc{HIERA}, we conduct experiments to answer the following
research questions.

\begin{rqbox}[Research Questions]
\textbf{RQ1: How does \textsc{HIERA} compare with existing methods in GPU kernel generation and optimization?}

\textbf{RQ2: How do implementation-space design choices and the key components of \textsc{HIERA} affect implementation validity and optimization performance?}

\textbf{RQ3: Can \textsc{HIERA} effectively be applied to other application scenarios, such as representative stencil computations in scientific computing?}

\end{rqbox}

\subsection{Benchmark and Task Preparation}

We evaluate \textsc{HIERA} on the first three levels of KernelBench~\cite{ouyang2025kernelbench}, comprising 250 PyTorch workloads: 100 Level-1 operator-level tasks, 100 Level-2 fused-operator tasks, and 50 Level-3 model-level workloads. These levels span individual tensor operators, multi-operator compositions, and complete model architectures. For all methods, we use the original PyTorch reference implementations and prescribed input-generation procedures, and evaluate the same task set under identical hardware and evaluation settings.

\subsection{Compared Methods and Experimental Variants}

\paragraph{Baselines.} We compare \textsc{HIERA} with three representative baselines: KernelBench-Caesar, CUDAForge, and CUDA-L1~\cite{ouyang2025kernelbench,zhang2025cudaforge,li2026cudal1improvingcudaoptimization}. KernelBench-Caesar is the official iterative workflow released with KernelBench, which generates and refines candidate implementations using execution feedback~\cite{ouyang2025kernelbench,kernelbench_repo}. CUDAForge is evaluated using its official Coder--Judge refinement pipeline~\cite{zhang2025cudaforge}. CUDA-L1 is a specialized training-based method evaluated using its official reproduction protocol~\cite{li2026cudal1improvingcudaoptimization}.

\paragraph{Base LLMs.} We use three base LLMs: DeepSeek-V3.2, Qwen3.6-Plus, and Gemini-3.6-Flash.

\paragraph{Implementation-space comparison.} To evaluate the effectiveness of adaptive cross-granularity planning, we compare \textsc{HIERA} with three fixed implementation-space variants on 90 randomly sampled KernelBench tasks, including 30 tasks from each of Levels~1--3. \textit{Pure CUDA} permits only custom CUDA implementations, \textit{CUDA Libraries} additionally allows optimized CUDA libraries, \textit{CUDA Libraries + PyTorch} further allows high-level PyTorch operators, \textsc{HIERA} dynamically selects the permitted implementation space according to workload characteristics and iterative feedback. All variants use the same sampled tasks, contract-augmented task specification, Qwen3.6-Plus LLM, optimization-direction planning procedure, candidate-generation budget, and evaluation protocol.

\paragraph{Ablation variants.} We construct two ablated variants to quantify the contributions of the principal components of \textsc{HIERA}. \textsc{HIERA} w/o Contract uses only the original task specification and requires the model to generate both the candidate implementation and all supporting files that are fixed in the contract-augmented specification. \textsc{HIERA} w/o Planning removes the hierarchical planning agent and directly refines candidate implementations using a fixed optimization prompt that imposes no constraints on the implementation space. Both variants use Qwen3.6-Plus and follow the same candidate-generation budget, verification procedure, and evaluation protocol as the complete framework.

\subsection{Implementation and Evaluation Protocol}

\paragraph{Main comparison protocol.} \textsc{HIERA}, KernelBench-Caesar, and CUDAForge each perform three refinement rounds with six candidates per round, yielding a maximum budget of \(B=18\) candidates per task. For each base LLM, we align the model version, sampling temperature of 0.3, candidate-generation budget, and evaluation settings while preserving the original candidate-selection and refinement mechanisms of each framework. Each candidate is compiled using the prescribed KernelBench task harness and verified against the original PyTorch reference implementation. Compilation or functional-verification failures are treated as invalid. Valid candidates are evaluated in FP32 using three warmup runs followed by 100 measurement runs, and the fastest valid implementation discovered within the full budget is used for comparison.

\paragraph{Limited-budget evaluation.} To evaluate optimization performance under limited search budgets, we report cumulative results at \(B\in\{1,6,12,18\}\). For the iterative methods, \(B=6\), \(12\), and \(18\) correspond to the cumulative candidates generated after the first, second, and third rounds, respectively, while \(B=1\) uses the first generated candidate. At each budget, we report the best valid implementation discovered among the first \(B\) candidates. All budget points are extracted from the same generation trajectories rather than from independently rerun experiments.

\paragraph{Case-study protocol.} We evaluate \textsc{HIERA} on a representative 2D box stencil from scientific computing. The stencil has radius \(R=3\), corresponding to a \(7\times7\) neighborhood with 49 stencil points. The input size is $10240 \times 10240$, and the stencil computation is repeated 10,240 times. We report the amortized per-step latency. The search consists of five iterations with a population size of ten, yielding a total budget of 50 candidate evaluations. We use cuDNN
\texttt{cudnnConvolutionForward} as the dense reference: a single-channel
$7\times7$ zero-padded cross-correlation (padding $3$, stride $1$,
FP64, \texttt{IMPLICIT\_PRECOMP\_GEMM}) implementing the same stencil.

\paragraph{Hardware and software environment.} Each optimization task is assigned to a single GPU. Table~\ref{tab:environment} summarizes the hardware and software environment.

\begin{table}[t]
\centering
\caption{Hardware and software environment.}
\label{tab:environment}
\begin{tabular}{ll}
\toprule
\textbf{Component} & \textbf{Configuration} \\
\midrule
GPU & NVIDIA A100-PCIE-40GB \\
CPU & Intel Xeon Gold 6430 \\
Operating system & Ubuntu 20.04.6 LTS \\
NVIDIA driver & 565.57.01 \\
CUDA Toolkit & 12.8 \\
Python & 3.12.13 \\
PyTorch & 2.7.1+cu126 \\
cuDNN & 9.5.1 \\
\bottomrule
\end{tabular}
\end{table}

\subsection{Evaluation Metrics}

We adopt the standard KernelBench evaluation metrics. Let \(s_i(B)\) denote the speedup of the best valid implementation discovered for task \(i\) within budget \(B\), relative to its PyTorch reference implementation. We report \(\mathrm{fast}_0\), the fraction of tasks with at least one functionally valid implementation; \(\mathrm{fast}_1\), the fraction whose best valid implementation outperforms the reference; and \(\mathrm{fast}_2\), the fraction whose best valid implementation achieves more than \(2\times\) speedup. 

\subsection{Scope and Limitations}

Our evaluation is subject to several scope constraints. All experiments are conducted on NVIDIA A100 GPUs. KernelBench is evaluated in FP32, whereas the stencil case study uses FP64; performance across other GPU architectures, multi-GPU settings, and precision formats remains to be investigated. Although latency measurements are repeated, candidate generation remains stochastic. Besides, considering the limited time budget, the stencil study only considers one operator and configuration and should therefore be viewed as evidence of feasibility rather than comprehensive validation across scientific computing workloads.

\section{Results}

\subsection{RQ1: Main Results}
\label{sec:rq1}

\subsubsection{Performance under limited candidate-generation budgets.}

\begin{figure}[t]
  \centering
  \includegraphics[width=\columnwidth]{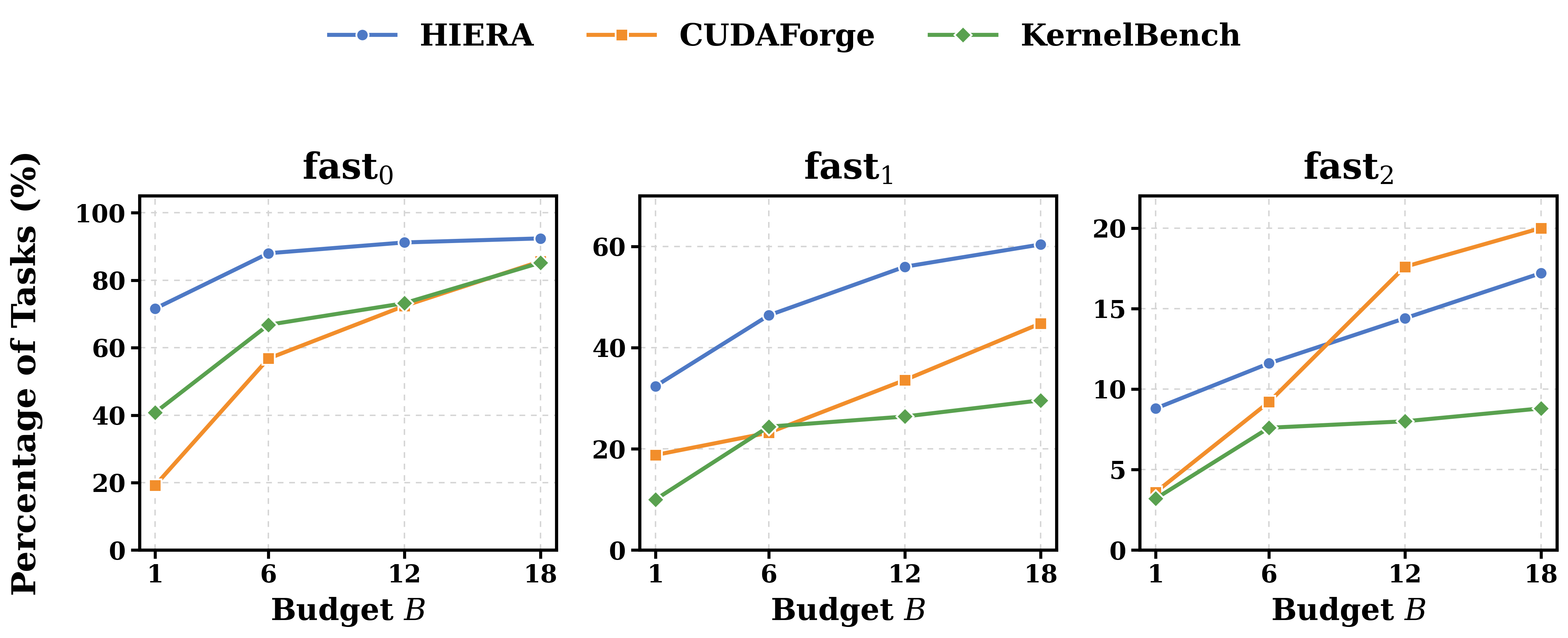}
  \caption{Limited-budget performance on KernelBench. At each budget $B$, we report the best valid implementation among the first $B$ generated candidates. All points are extracted from the same generation trajectories.}
  \label{fig:limited-budget}
\end{figure}

Figure~\ref{fig:limited-budget} shows that HIERA achieves substantially stronger
sample efficiency and optimization quality across search budgets. At $B=1$, it achieves $71.6\%$ fast$_0$ and $32.4\%$ fast$_1$, outperforming KernelBench by $30.8$ and $22.4$ percentage points and CUDAForge by $52.4$ and $13.6$ points, respectively. This advantage persists at $B=18$: HIERA reaches $92.4\%$ fast$_0$ and $60.4\%$ fast$_1$, exceeding KernelBench ($85.2\%$/ $29.6\%$) and CUDAForge ($85.6\%$/ $44.8\%$).  For the more stringent fast$_2$ metric, HIERA also leads at small budgets ($8.8\%$ at $B=1$ and $11.6\%$ at $B=6$). Overall, HIERA provides the best budget--quality trade-off for correctness and acceleration.

\paragraph{Comparison with baselines and base LLMs.}

Table~\ref{tab:main_comparison} compares \textsc{HIERA} with KernelBench and CUDAForge under three base LLMs, together with the training-based CUDA-L1 baseline. We report $\mathrm{fast}_0$, $\mathrm{fast}_1$, and $\mathrm{fast}_2$ following the KernelBench protocol. Higher values are better.

\begin{table*}[t]
\centering
\caption{Comparison with baselines under different base LLMs. Results are percentages. Bold indicates the best result among methods using the same base LLM. $\dagger$ indicates that CUDA-L1 does not vary with the base LLM setting and is therefore reported once.}
\label{tab:main_comparison}
\small
\setlength{\tabcolsep}{5.41pt}
\renewcommand{\arraystretch}{1.13}
\begin{tabular}{llccc|ccc|ccc}
\toprule
\textbf{Base LLM} & \textbf{Method} & \multicolumn{3}{c|}{\textbf{Level 1}} & \multicolumn{3}{c|}{\textbf{Level 2}} & \multicolumn{3}{c}{\textbf{Level 3}} \\
& & $\mathrm{fast}_0$ & $\mathrm{fast}_1$ & $\mathrm{fast}_2$ & $\mathrm{fast}_0$ & $\mathrm{fast}_1$ & $\mathrm{fast}_2$ & $\mathrm{fast}_0$ & $\mathrm{fast}_1$ & $\mathrm{fast}_2$ \\
\midrule
\multirow{3}{*}{DeepSeek-V3.2}
& \textsc{HIERA} & \textbf{91} & \textbf{56} & \textbf{22} & \textbf{90} & \textbf{35} & 9 & \textbf{64} & \textbf{42} & \textbf{10} \\
& KernelBench-Caesar & 77 & 17 & 3 & 80 & 34 & 5 & 46 & 12 & 2 \\
& CUDAForge & 90 & 22 & 6 & 78 & 27 & \textbf{15} & 60 & 32 & \textbf{10} \\
\midrule
\multirow{3}{*}{Qwen3.6-Plus}
& \textsc{HIERA} & \textbf{97} & \textbf{68} & \textbf{18} & \textbf{99} & \textbf{62} & 20 & 70 & \textbf{42} & \textbf{10} \\
& KernelBench-Caesar & 91 & 21 & 6 & 90 & 40 & 13 & 64 & 26 & 6 \\
& CUDAForge & 90 & 47 & \textbf{18} & 88 & 45 & \textbf{28} & \textbf{72} & 40 & 8 \\
\midrule
\multirow{3}{*}{Gemini-3.6-Flash}
& \textsc{HIERA} & \textbf{95} & \textbf{63} & \textbf{20} & \textbf{96} & \textbf{50} & 14 & 68 & \textbf{40} & \textbf{12} \\
& KernelBench-Caesar & 80 & 19 & 4 & 85 & 32 & 11 & 52 & 14 & 2 \\
& CUDAForge & 92 & 45 & 16 & 91 & 40 & \textbf{20} & \textbf{70} & 34 & 8 \\
\midrule
-- & CUDA-L1$^\dagger$ & 74 & 19 & 10 & 81 & 36 & 13 & 72 & 50 & 6 \\
\bottomrule
\end{tabular}
\end{table*}

\textsc{HIERA} consistently delivers the strongest optimization quality across diverse base LLMs, achieving the best or tied-best result in 22 of 27 comparisons against the LLM-conditioned baselines. Its most pronounced advantage is in $\mathrm{fast}_1$, where it ranks first among the inference-time methods across all workload levels and base models, showing that \textsc{HIERA} reliably converts correct generations into practical speedups rather than merely improving implementation validity. It also leads $\mathrm{fast}_0$ in seven of nine level--model settings, reaching 99\% on Qwen3.6-Plus Level~2 and 96\% on Gemini-3.6-Flash Level~2. Despite requiring no reinforcement-learning training, \textsc{HIERA} outperforms the training-based CUDA-L1 on all Level~1 and Level~2 metrics when averaged across the three base LLMs.

\subsection{RQ2: Design Effectiveness}
\label{sec:rq2}

\subsubsection{Cross-Granularity Search-Space Comparison}

\begin{figure}[t]
  \centering
  \includegraphics[width=\columnwidth]{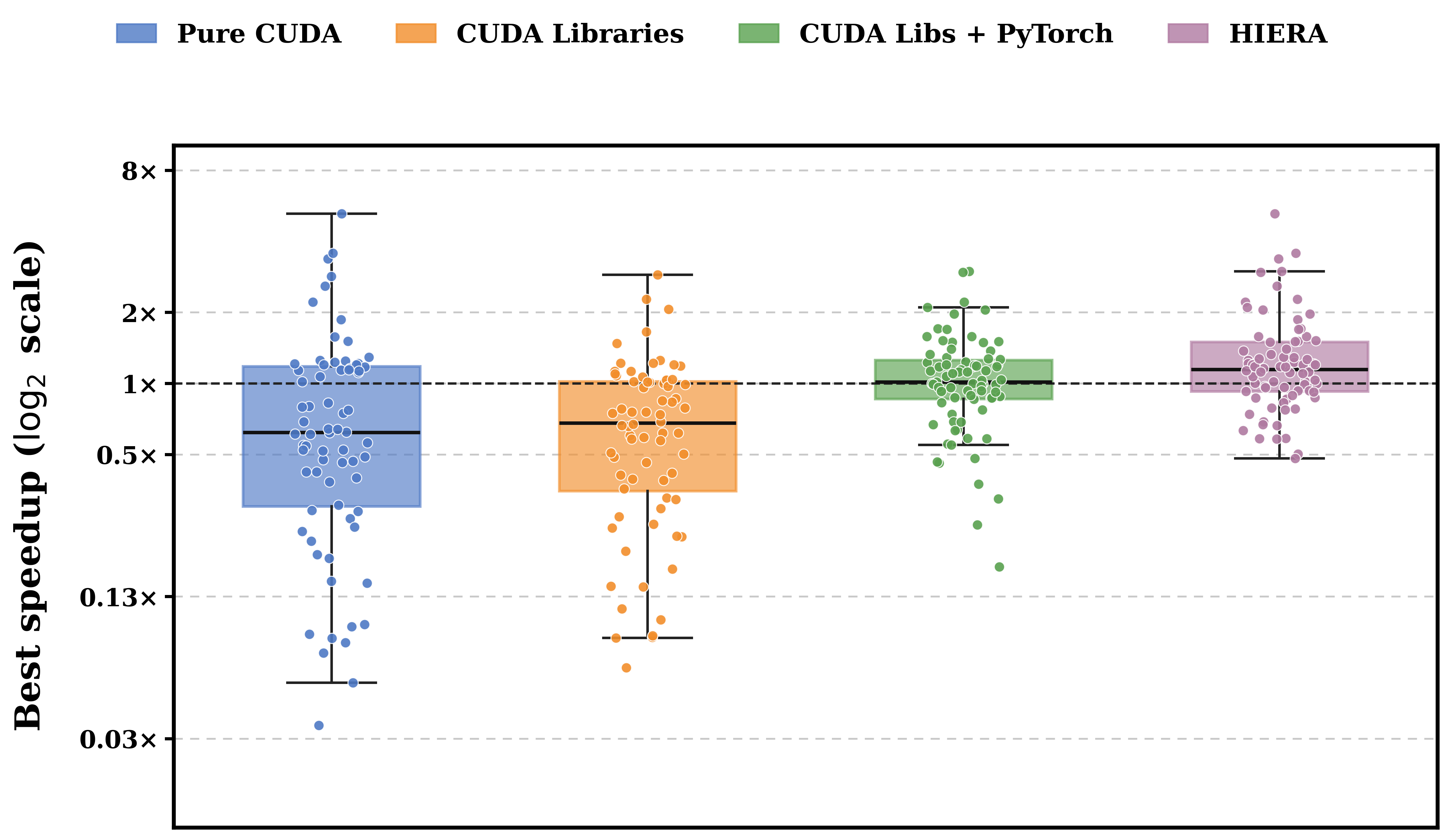}
 \caption{Distribution of the best verified speedup achieved by three fixed implementation spaces and \textsc{HIERA} on 90 KernelBench tasks sampled from Levels~1--3. }
  \label{fig:search_space}
\end{figure}

Figure~\ref{fig:search_space} shows that implementation-space selection substantially affects both optimization quality and result stability. \textit{Pure CUDA} offers the greatest optimization potential, reaching a maximum speedup of $9.32\times$, but yields a mean of only $1.00\times$, a median of $0.62\times$, and the largest variance ($1.79$). Expanding the space to \textit{CUDA Libraries} reduces the variance to $0.27$, while \textit{CUDA Libraries + PyTorch} further improves the mean and median to $1.10\times$ and $1.01\times$, respectively, with the lowest variance ($0.26$), although its observed maximum speedup is $2.98\times$. By adaptively selecting the implementation space, \textsc{HIERA} achieves the highest mean ($1.42\times$) and median ($1.15\times$), while reducing variance by $23.4\%$ relative to \textit{Pure CUDA}. These results show that \textsc{HIERA} combines the stability of broader implementation spaces with the optimization potential of custom CUDA.

\subsubsection{Ablation Study}

\begin{figure*}[t]
\centering
\begin{minipage}[t]{0.495\textwidth}
\centering
\includegraphics[width=\linewidth]{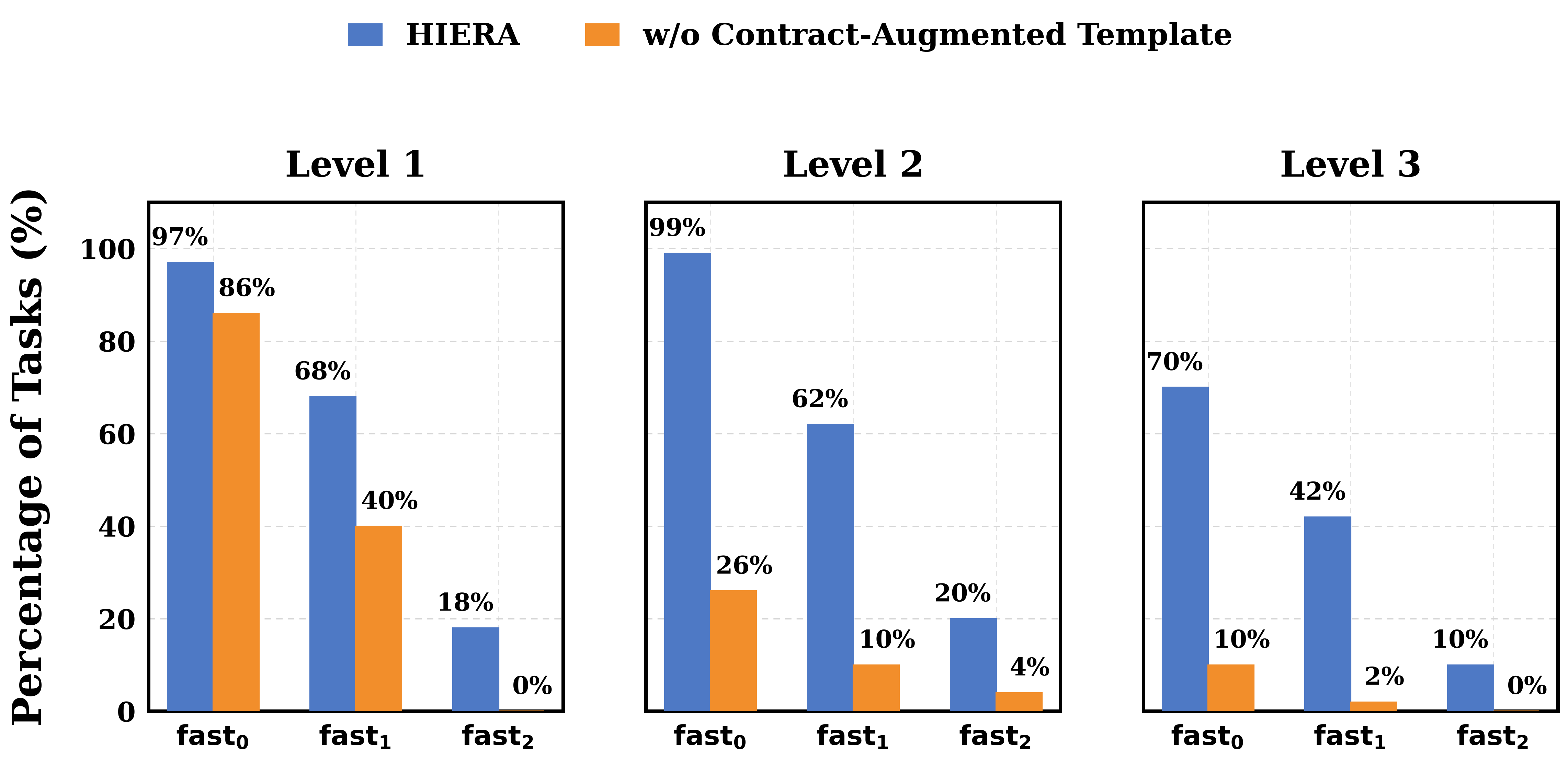}
\vspace{-2mm}
\small (a) Effect of removing the contract-augmented template layer.
\end{minipage}
\hfill
\begin{minipage}[t]{0.495\textwidth}
\centering
\includegraphics[width=\linewidth]{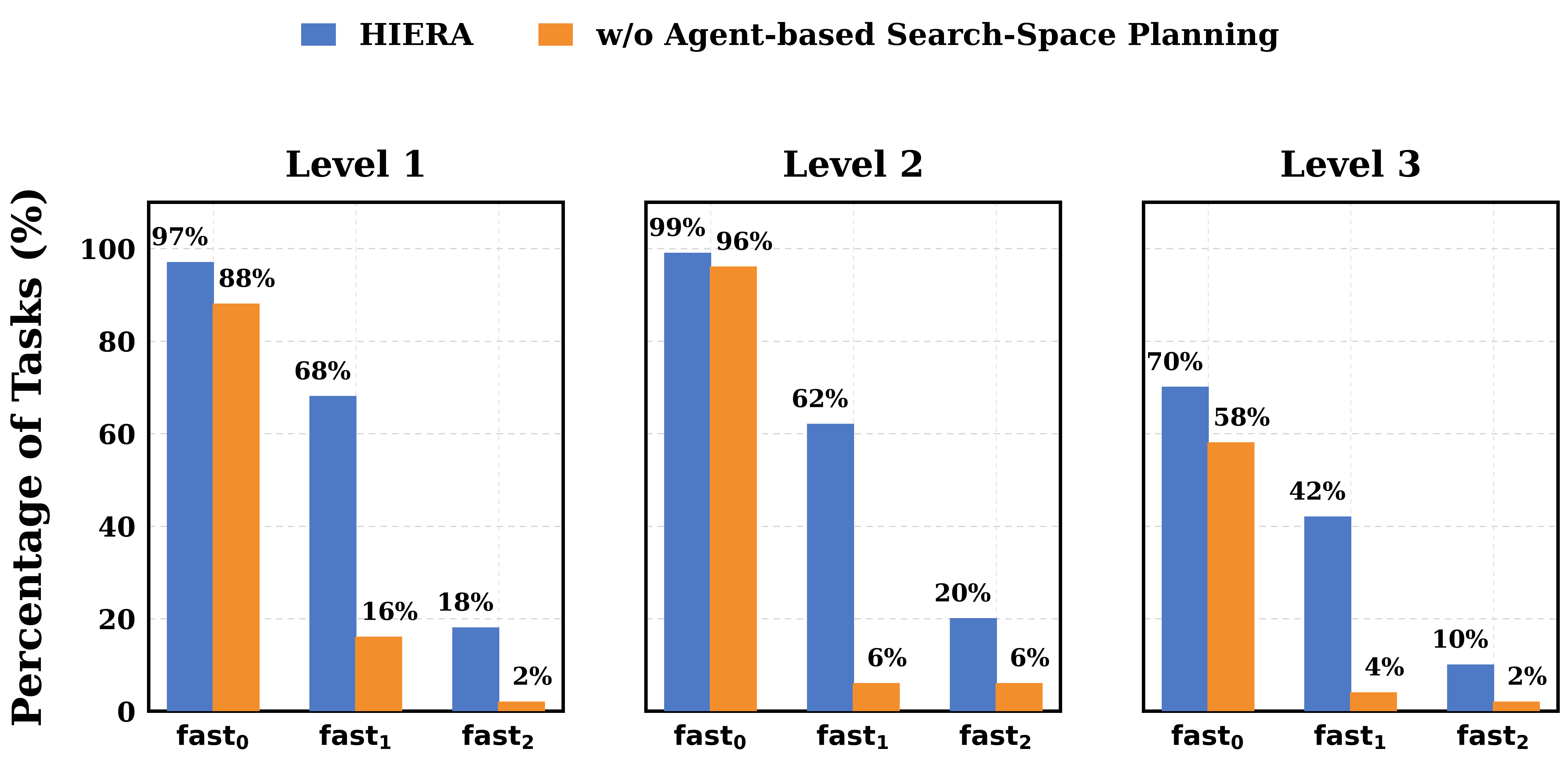}
\vspace{-2mm}
\small (b) Effect of removing Agent-based search-space planning.
\end{minipage}
\caption{Ablation results across KernelBench Levels 1--3. Blue bars denote the complete \textsc{HIERA} framework. Orange bars denote the corresponding ablated variant.}
\label{fig:ablation}
\end{figure*}

\paragraph{Effect of contract-augmented templates.} Removing the contract-augmented specification consistently degrades both implementation validity and optimization performance, with the effect increasing with workload complexity. On Level~1, $\mathrm{fast}_0/\mathrm{fast}_1/\mathrm{fast}_2$ decrease from $97/68/18\%$ to $86/40/0\%$. The degradation is substantially larger on Levels~2 and~3, where $\mathrm{fast}_0$ drops by 73 and 60 percentage points, respectively. These results show that explicit interfaces, parameter semantics, and fixed supporting artifacts are increasingly important for producing valid implementations on complex workloads.

\paragraph{Effect of hierarchical search-space planning.} Removing hierarchical planning has a smaller effect on validity but sharply reduces acceleration rates. On Level~2, $\mathrm{fast}_0$ decreases only from $99\%$ to $96\%$, whereas $\mathrm{fast}_1$ and $\mathrm{fast}_2$ fall from $62/20\%$ to $6/6\%$. Similar reductions occur on Levels~1 and~3, where $\mathrm{fast}_1$ drops by 52 and 38 percentage points. Thus, generic refinement can still produce valid implementations, but rarely identifies implementation spaces and optimization directions that yield practical speedups.

In summary, contract augmentation primarily preserves feasibility, whereas hierarchical planning improves acceleration; their complementary effects enable reliable optimization across workload levels.

\subsection{RQ3: Case Study on Stencil Computation}
\label{sec:rq3}

\begin{figure}[t]
    \centering
    \includegraphics[width=0.92\linewidth]{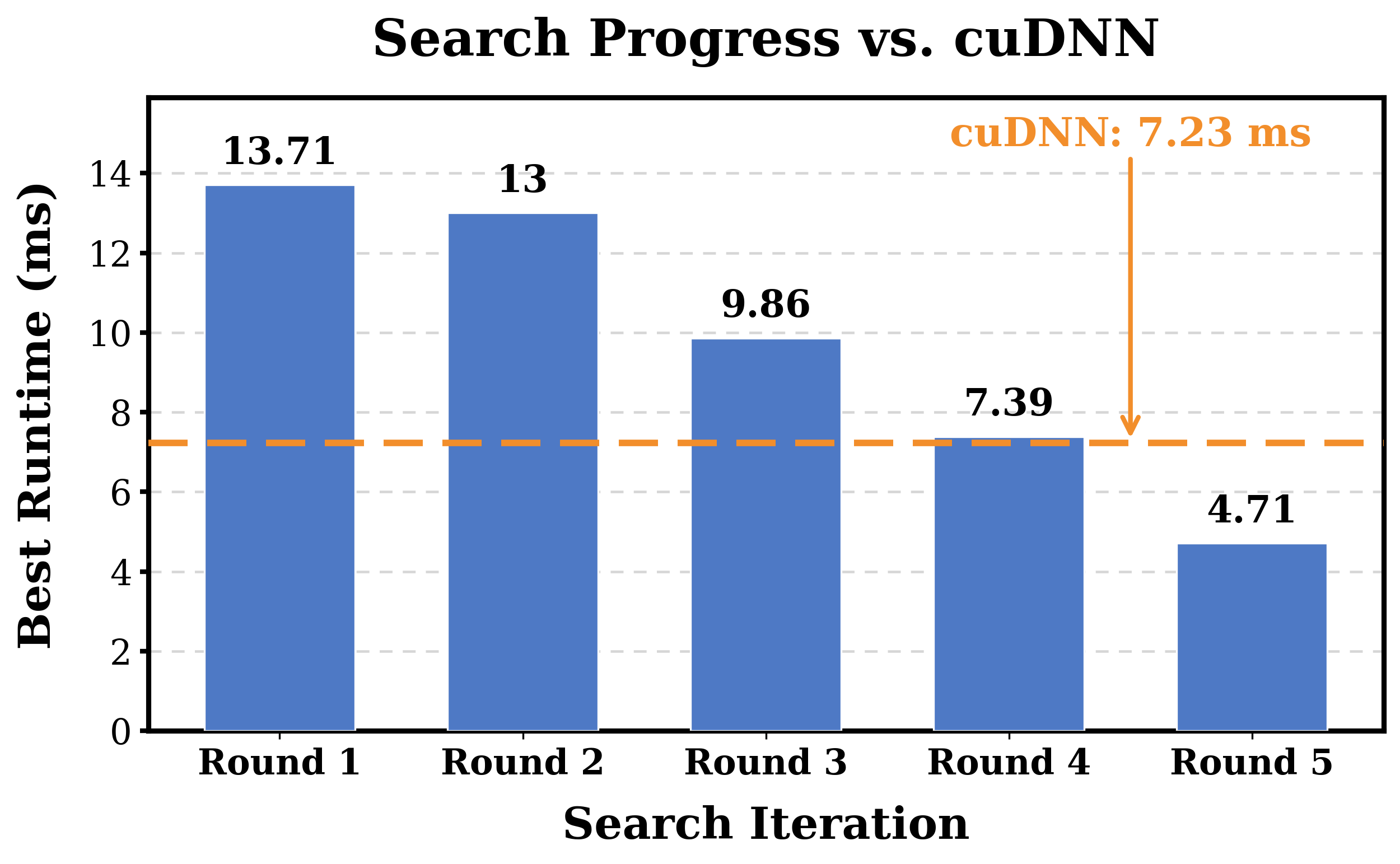}
    \caption{
    Search progress compared with cuDNN on the 2D box stencil workload
    ($R=3$, 49 stencil points). 
    }
    \label{fig:stencil-search-vs-cudnn}
\end{figure}

Figure~\ref{fig:stencil-search-vs-cudnn} shows that the runtime decreases from $13.71$\,ms in the first iteration to $4.71$\,ms in the fifth iteration. The first four iterations remain slower than or close to the cuDNN reference of $7.23$\,ms; the fourth iteration reaches $7.39$\,ms. In the fifth iteration, the search discovers a candidate with a runtime of $4.71$\,ms, which is $34.8\%$ lower than the cuDNN runtime and corresponds to a $1.53\times$ speedup. Overall, the search reduces runtime by $65.6\%$ relative to the first-round candidate, demonstrating that the limited budget is sufficient to find an implementation that outperforms cuDNN.

\section{Conclusion}

We presented \textsc{HIERA}, a workload-aware framework that explicitly plans implementation spaces and optimization directions for GPU kernel optimization. By combining contract-augmented task specifications with feedback-driven refinement, \textsc{HIERA} balances implementation reliability and optimization flexibility under limited search budgets. Future work will extend the framework across GPU architectures, precision formats, and implementation backends, and explore self-improving planning from accumulated optimization trajectories.

\bibliography{hiera_references_aaai27}


\end{document}